\documentclass[oldversion]{aa}     
\def \th {\thinspace}

\def\approxgt{\mathrel{\hbox{\rlap{\lower.55ex \hbox {$\sim$}} \kern-.3em \raise.4ex \hbox{$>$}}}}
\def\lesssim{\mathrel{\hbox{\rlap{\lower.55ex \hbox {$\sim$}} \kern-.3em \raise.4ex \hbox{$<$}}}}
\def\approxlt{\mathrel{\hbox{\rlap{\lower.55ex \hbox {$\sim$}} \kern-.3em \raise.4ex \hbox{$<$}}}}
\def \degmark {^\circ}
\def \sun {\hbox {$\odot$}}
\usepackage{epsf}
\usepackage{graphicx}
\begin{document}

\title{An explanation of the Z-track sources}

\author{ M. J. Church\inst{1,2}
\and G. S. Halai\inst{1}
\and M. Ba\l uci\'nska-Church\inst{1,2}}
\institute{School of Physics and Astronomy, University of Birmingham,
           Birmingham, B15 2TT, UK\\
\and
          Astronomical Observatory, Jagiellonian University, 
          ul. Orla 171, 30-244 Cracow, Poland.\\}
\offprints{mjc@star.sr.bham.ac.uk}

\date{Received 20 February 2006; Accepted 4 September 2006}
\titlerunning{Explaining the Z-track sources}
\authorrunning{Church et al.}



\abstract{We present an explanation of the Z-track phenomenon based on spectral fitting results
of {\it Rossi-XTE} observations of the source GX\th 340+0 using the emission model
previously shown to describe the dipping Low Mass X-ray Binaries. In our Z-track model,
the soft apex is a quiescent state of the source with lowest luminosity. Moving away
from this point by ascending the normal branch the strongly increasing luminosity of the 
Accretion Disc Corona (ADC)
Comptonized emission $L_{\rm ADC}$ provides substantial evidence for a large increase
of mass accretion rate $\dot M$. There are major changes in the neutron star blackbody 
emission, $kT$ increasing to high values, the blackbody radius $R_{\rm BB}$ decreasing, 
these changes continuing monotonically on both normal and horizontal branches. 
The blackbody flux increases by a 
factor of ten to three times the Eddington flux so that the physics of the horizontal 
branch is 
dominated by the high radiation pressure of the neutron star, which we propose disrupts 
the inner disc, and an increase of column density is detected. We further propose that the 
very strong radiation pressure is responsible for the launching of the jets detected in 
radio on the horizontal branch. On the flaring branch, we find that $L_{\rm ADC}$ is
constant, suggesting no change in $\dot M$ so that flaring must consist of unstable
nuclear burning. At the soft apex, the mass accretion rate per unit area on the neutron 
star $\dot m$ is minimum for the horizontal and normal branches and about equal to 
the theoretical upper limit for unstable 
burning. Thus it is possible that unstable burning begins as soon as the source arrives
at this position, the onset of unstable burning being consistent with theory. The large 
increase in $R_{\rm BB}$ in flaring is reminiscent of radius expansion in X-ray bursts. 
Finally, in our model, $\dot M$ does not increase monotonically along the Z-track as 
often previously thought.
\keywords{Accretion: accretion discs -- acceleration of particles -- binaries: 
close -- line: formation -- stars: neutron -- X-rays: 
binaries -- X-rays: individual (GX\th 340+0)}}
\maketitle

\section{Introduction}

The Z-track sources form the brightest group of Low Mass X-ray Binary (LMXB)
sources containing a neutron star, consisting of 6 members: 
GX\th 340+0, GX\th 5-1, Cyg\th X-2, \hbox{Sco\th X-1}, GX\th 17+2 and GX\th 349+2
(Hasinger \& van der Klis 1989).
All of these persistently radiate at about the Eddington 
limit for a neutron star and do not, in general, exhibit X-ray bursting characteristic
of LMXBs with lower luminosities. The sources trace a Z-shaped pattern in X-ray 
hardness versus intensity or hardness versus softness demonstrating strong spectral
evolution associated with major physical changes within the sources.
However, the nature of these changes is not understood 
and this remains a significant astrophysical problem.
Moreover, the Z-track sources are all detected as radio emitters 
showing relativistic jets to be present, but essentially on one branch only
(e.g. Penninx 1989). The mechanism
of jet formation also remains an important astrophysical problem. The occurrence of
jets on one branch only may allow us to probe how jets are launched.

Hasinger et al. (1989) showed that the sources could be found on   
three different branches of a skewed Z-shape in a hardness-intensity diagram: 
the horizontal branch (HB), the normal branch (NB) and the flaring branch (FB). 
These diagrams demonstrated in
a model-independent way that strong spectral evolution was taking place around
the Z-track, suggestive of changes at the compact object and accretion disc,
but did not reveal the nature of the changes. It was found 
that quasi-periodic oscillations (QPO) exhibit systematic changes
around the Z-track (e.g. van der Klis et al. 1987; Hasinger \& van der Klis 1989). 
Clearly, investigation of the spectral evolution along the Z-track
is likely to reveal the nature of 
the changes; however, spectral studies of the Z-track sources have been 
hindered by a lack of agreement over which emission model for LMXBs to use. 

\subsection{Spectral fitting}
The nature of X-ray emission in LMXBs has been controversial with two main types of
model for the continuum: the first is the Eastern model comprising
multi-colour blackbody emission from the inner accretion disc plus
non-thermal emission from a central Comptonizing region (Mitsuda et al. 1989);
the second model discussed below comprises neutron star blackbody emission plus
Comptonization in an extended corona. Previous spectral fitting of the Z-track sources 
has been largely based on the Eastern model.
Spectral evolution along the Z-track in several sources was investigated by
Schulz and Wijers (1993) and Schulz et al. (1989). 
More recently,
spectral fitting of the source Cyg\th X-2 was carried out by Done et al. 
(2002) 
using the Eastern model.
Spectral fitting results were given in terms of the parameters of the disc blackbody and thermal 
Comptonization emission components, but it was not clear what caused the parameters to change 
in the way found. The same model was used by Agrawal \& Sreekumar (2003) to fit {\it Rossi-XTE}
spectra of GX\th 349+2.
Spectral fitting of broadband {\it BeppoSAX} data was carried out for GX\th 17+2, 
GX\th 349+2 and for Cyg\th X-2 (di Salvo et al. 2000, 20001, 2002)
who argued that all three sources could be fitted by the Eastern model.
In GX\th 17+2 and GX\th 349+2, there was little change in blackbody parameters 
with position on the Z-track, however, in Cyg\th X-2 there was an increase of blackbody 
temperature and decrease of blackbody radius moving from the horizontal to the normal
branch, suggesting that the inner disc radius was shrinking although an explanation
of this was not proposed.


Theoretical models for the Z-track sources have been proposed (Psaltis et al. 1995), 
involving a magnetosphere
at the inner accretion disc, and changes in the geometry and extent of the magnetosphere.
However, it has not been possible to test this model in detail, as, for example, the geometry
cannot be determined. This modelling also assumes the main element of the
Eastern model, that the Comptonizing region is a small central region, and evidence
(below) does not support this.

\subsection{Approach of the present work}

Thus previous spectral fitting based on the Eastern model has not produced
a clear explanation of the Z-track and there is now substantial evidence against this model
provided by the dipping class of LMXBs.
In these sources with high inclination (65 - 85$\degmark$), absorption of
the X-ray emission takes place on every orbital cycle in the
bulge in the outer disc. Complex spectral evolution in dipping strongly 
constrains emission models providing substantial evidence
for a model comprising simple, unmodified blackbody emission from the neutron star and Comptonized emission
from an extended accretion disc corona of radial extent $R$ 
typically 50000 km, and height $H$ having \hbox{$H/R$ $<$ 1} (the ``Birmingham model''; 
Church \& Ba\l uci\'nska-Church 1995, 2004). The extended size of
the ADC is demonstrated 
%
by dip ingress timing (Church \& Ba\l uci\'nska-Church 2004)
which allows measurement of the ADC radial extent. The spectra of the dipping LMXBs in the
non-dip state and in every stage of dipping from many observations have been well-fitted
using this model (Church et al. 1997, 1998a,b, 2005; Ba\l uci\'nska-Church et al. 1999, 2000;
Smale et al. 2001; Barnard et al. 2001), and so the evidence for the model is substantial
and hence argues against the Eastern model with a small central Comptonizing region.

It should not be thought that the spectral model used is {\it ad hoc}, i.e. having a simple form easily implemented
that happens to fit the spectra, 
but with possibly large inaccuracies. The form used is {\sc bb + cpl}, where {\sc bb} is simple blackbody emission
from the neutron star and {\sc cpl} is a cut-off power law representing Comptonization.
The correct form for Comptonization is very dependent on
the correct specification of the seed photon spectrum. With an extended ADC, the seed photons must 
originate in the disc below the corona with a large population of soft photons below 0.1 keV from large 
radii. We previously showed that a power law or cut-off power law represents Comptonization well
allowing for this, and the Comptonized spectrum continues to rise below 1 keV (Church \& Ba\l uci\'nska-Church 
2004) whereas the {\sc comptt} model often used incorporates a blackbody seed photon spectrum 
decreasing strongly below 1 keV. We have never been able to detect any departure of the thermal
emission from simple blackbody form, and we previously reviewed the evidence for use of
a simple blackbody and discussed in detail how possible modification
in the neutron star atmosphere depends critically on the electron density which is very poorly known 
(Ba\l uci\'nska-Church et al. 2001). Kuulkers et al. (2002) also present evidence that X-ray burst emission
is of simple blackbody shape.

Finally, when the spectra of LMXBs are fitted with the Eastern model in the form disc blackbody + cut-off
power law (e.g. Church \& Ba\l uci\'nska-Church 2001), 
it is often found that the inner radius of the disc blackbody is unphysically small with values
less than 1 km, i.e. substantially less than the neutron star radius. Done et al. (2002) argued that
this was purely an artefact of using an incorrect form (cut-off power law) for the Comptonized emission.
The above argument shows that, on the contrary, the cut-off power law
is a proper description of the Comptonized spectrum if the ADC is extended
(at least below the high energy cut-off) as discussed fully in Church \& Ba\l uci\'nska-Church (2004). Thus the
conclusion that the Eastern model gives unphysically small inner disc radii is valid, and in addition,
modelling based on assuming seed photons that are blackbody with $kT$ $\sim$ 1 keV as in {\sc comptt}
is invalid.

Thus the approach used in the present work consists of applying for the first time the Birmingham 
emission model to the Z-track sources, as it provided a good explanation of the dipping LMXBs, to
determine whether this model suggests an explanation of the Z-track phenomenon.
We chose the source GX\th 340+0 since high quality {\it Rossi-XTE}
data extending round the full Z-track were available. The source was classified as a Z-track 
source by Hasinger \& van der Klis (1989). QPO were found in {\it Exosat} data (van Paradijs 
et al. 1988) and in {\it Ginga} data (Penninx et al. 1991) and kilohertz QPO were discovered 
by Jonker et al. (1998). The detection of radio by Penninx et al. (1993) is important
indicating the presence of jets. Fender \& Hendry (2000)
summarized previous work on radio detection from X-ray binaries showing that radio emission
is detected from black hole binaries and from the Z-track sources 
when on the Horizontal Branch,
concluding that a requirement for jet formation was an X-ray luminosity approaching the
Eddington limit, i.e. greater than 0.1 $L_{\rm Edd}$. More recently, Migliari \& Fender (2006)
summarize the detection of relatively weak radio emission from a number of Atoll sources.

\subsection{A monotonic increase of mass accretion rate ?}
Finally, we address a widely-held view that the observed hardness changes in the Z-track
sources are driven in some way by a mass accretion rate $\dot M$ which increases 
monotonically around the Z-track in the direction HB-NB-FB.
The movement round the track
on timescales of hours, without jumping between tracks, suggested the variation of a
single parameter, probably $\dot M$ (Priedhorsky et al. 1986). 
In a multi-wavelength campaign (Hasinger et al. 1990), \hbox{Cyg\th X-2} was observed in
X-ray, ultraviolet, optical and radio. The results of the {\it IUE} observations (Vrtilek et al. 1990)
showed an increase of intensity for motion HB-NB-FB. However, the identification of these data 
with the flaring branch depended on strong variability which was probably due to
X-ray dipping, and omitting these data would mean that there is no very dramatic
correlation of UV with track position. Van Paradijs et al. (1990) presented evidence for
increase of optical emission line strength for motion HB-FB but noted that the number of data
points was rather limited, while absorption line strength decreased, indicating that the processes 
involved are complex and can involve X-ray reprocessing.
\begin{figure*}                                                           
\includegraphics[width=64mm,height=176mm,angle=270]{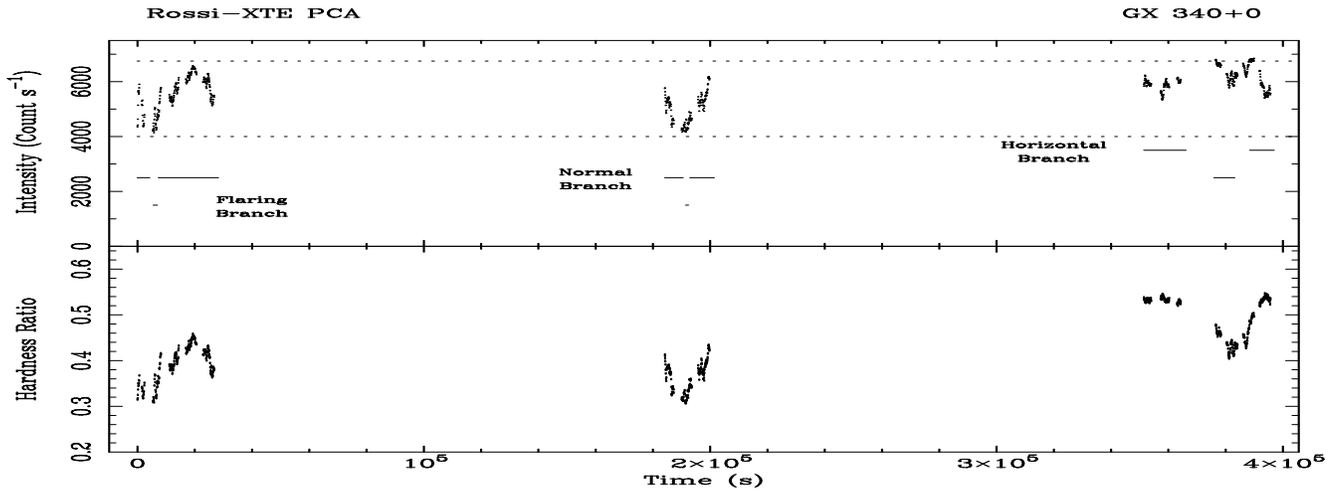}
\caption{Top: Background-subtracted and deadtime-corrected PCA light curve of the 1997 September observation
of GX\th 340+0 with 64 s binning. Parts of the lightcurve are identified (see text) with branches of the 
Z-track (Fig. 2) and shown as horizontal bars. Most of the first half of the observation is on the Normal 
Branch except for two large flares as shown; the second half of the observation is mostly on the Horizontal Branch. The 
dotted lines emphasize the upper and lower intensity limits which are consistent across the observation: 
these limits correspond to the intensities of the Hard Apex and Soft Apex of the Z-track. Bottom: the
corresponding variation of hardness ratio (7.3 -- 18.1 keV)/(4.1 -- 7.3 keV).}
\label{}
\end{figure*}
%

A monotonic increase of $\dot M$ around the Z-track in the direction HB-NB-FB is, of course,
inconsistent with the basic observational fact that the X-ray intensity {\it decreases}
between the hard apex and soft apex. A monotonic increase of $\dot M$ is not an assumption
of the Eastern model; however Agrawal et al. (2003) suggested that the results of fitting this model
might be qualitatively consistent with a monotonic increase. It would be necessary for the 
Comptonized emission luminosity to decrease from hard to soft apex while $\dot M$ increased.
In the present work, the luminosity decrease by a factor of two seems unlikely 
if $\dot M$ were increasing.

Kuulkers et al. (2002) provided another argument against the increase of $\dot M$ around the Z-track.
Unstable nuclear burning is expected at various mass
accretion rates (Fujimoto et al. 1981; Fushiki \& Lamb 1987; Bildsten 1998; Schatz et al. 
1999) including 
accretion rates close to the Eddington $\dot M$ due to He burning in a mixed H/He 
environment, i.e. $1.0\times 10^{-9} <\dot M < 2.6\times 10^{-8}$ M$_{\sun}$ yr$^{-1}$.
(Bildsten 1998). At higher mass accretion rates corresponding to luminosities 
greater than $3\times 10^{38}$ erg s$^{-1}$, burning is expected to be stable.
Thus, Kuulkers et al. argued that  
unstable nuclear burning should be observed in the Z-track sources, and X-ray bursting 
is not, in general, seen. Bursts are seen in GX\th 17+2 and Cyg\th X-2 
(Kahn \& Grindlay 1984; Tawara et al. 1984; Sztajno et al. 1986; Kuulkers et al. 1995;
Wijnands et al. 1997; Smale 1998). However, bursting in these sources 
showed no correlation of burst properties with Z-track position (Kuulkers et al. 1995, 
1997) as expected if the mass accretion rate was changing substantially along
the Z-track. 
It thus appears that the increase of $\dot M$ around 
the Z-track cannot be regarded as proven, and the results presented
here will be viewed in this context.

\section{Observations and analysis}

Several observations have been made of GX\th 340+0 with the {\it Rossi X-ray Timing Explorer}
satellite. We examined the data in the {\sc heasarc} archive remotely obtaining hardness-intensity
plots for all observations of this source, and selected data covering a full Z-track
in a relatively short time (to prevent assembling data in which several Z-tracks occurred, shifted 
with respect to each other, which would confuse the analysis).
We analysed the long observation made in 1997 September, consisting of four
sub-observations each of between 5 and 8 hours duration and spanning a total time of 400 ksec, 
starting on September 21 and ending on September 25. Data from both the proportional counter
array (PCA) and the high-energy X-ray timing experiment (HEXTE) were used (full bands: 2 -- 60 keV
and 15 -- 250 keV, respectively). The PCA was 
in Standard2 mode with 16 s resolution. It consists of five Xe proportional counter 
units (PCU) with a combined effective area of about 6500 cm$^2$ (Jahoda et al. 1996). Examination 
of the housekeeping data showed that all 5 PCUs were reliably on during this observation. 
Light curves and spectra were extracted using the standard {\it RXTE} analysis software
{\sc ftools 5.3.1}. PCA lightcurves were extracted from the raw data
for the top layer of the detector as normal
using both left and right anodes. Standard screening was applied to select data with an offset 
between the source and telescope pointing direction of less than 0.02$\degmark$, and elevation 
above the Earth's limb of more than 10$\degmark$. A lightcurve was extracted in the energy 
band 1.9 -- 18.5 keV.
The program {\sc pcabackest} was used to generate background files corresponding to each
PCA raw data file and these were used for background subtraction. The latest background model
was applied, specifically the ``bright'' model recommended for Epoch 3 of the mission (defined
as 1996 April 15 -- 1999 March 22). Deadtime correction was carried out using dedicated
software encapsulating the prescription provided by the mission specialists. The 
background-subtracted, deadtime-corrected lightcurve is shown with 64 s binning 
in Fig. 1 (top panel).
PCA lightcurves were also extracted in three bands spanning the total band used
in Fig. 1: 1.9 -- 4.1 keV (low band), 4.1 -- 7.3 keV (medium band) and 7.3 -- 18.5 keV
(high band). From these, a hardness-intensity plot was constructed defining hardness
as the ratio of high/medium. This full Z-track is shown in Fig. 2.

Lightcurves and spectra were also extracted from Cluster 1 of the HEXTE instrument, 
so that simultaneous spectral fitting of PCA and HEXTE data could be carried out.
These were extracted with the {\sc ftool} 
{\sc hxtlcurv} which also provided background files and allowed deadtime correction 
based on the deadtime coefficients file of 2000 February.

\section{Results}

\subsection{The X-ray lightcurve and the Z-track}

\begin{figure}                                                       
\includegraphics[width=84mm,height=84mm,angle=270]{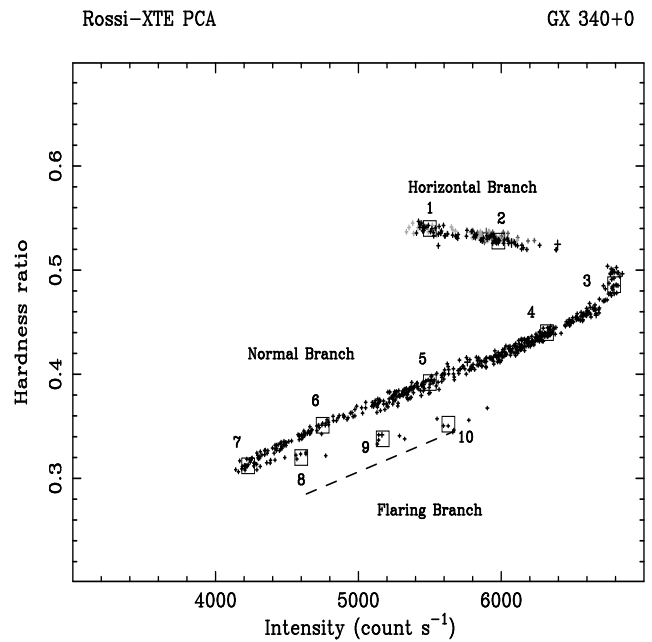}
\caption{Z-track of the observation derived from background-subtracted and deadtime-corrected lightcurves
with 64 s binning. Intensity is in the band 1.9 -- 18.1 keV and hardness ratio is (7.3 -- 18.1 keV)/(4.1 -- 7.3 keV).
The boxes show the ranges of intensity and hardness ratio used for selection of
PCA and HEXTE spectra, and are labelled 1 -- 10 starting from the Horizontal Branch end of the Z-track.
The dashed line shows the position of 7 data points removed before analysis (see text).}
\label{}
\end{figure}

Figure 1 shows the strong variability of GX\th 340+0 during these observations.
Firstly, an identification was made of each part of the PCA lightcurve with position
on the Z-track (Fig. 2). Individual short segments of data were selected from the lightcurve using the
{\sc ftool} {\sc maketime} and for each segment a hardness-intensity plot made
so as to reveal the Z-track position of each, and these identifications
are labelled in Fig. 1. The lower panel shows the hardness ratio (HR) as a function of time from which 
the positive correlation of HR and intensity on the normal branch can be seen, as well as
the reciprocal variation on the HB. The source spent most time in the horizontal and 
normal branches, and it can be seen that the source
intensity varied between two levels shown as dotted lines. At the start of the 
observation the source moved along the normal branch from the soft apex to the
hard apex, the intensity increasing from 4000 to 6800 count s$^{-1}$ per PCU.
In the central sub-observation, the source moved down the normal branch to lower intensities
and then back to higher intensities on this branch. In the third and fourth
sub-observation, the source was mostly on the horizontal branch; i.e. having an intensity
reduced from the upper limit found at the hard apex. Two short sections of data 
were on the flaring branch, each about 1000 s in duration, i.e. there were two flares;
however, the intensity of the source was so high that these provided more than adequate
counts for high quality spectral fitting (below). A section of data near the start of
the observations did not lie on any of the three normal branches,
but extended downwards from the extreme end of the flaring branch. Such behaviour
has previously been seen, for example in GX\th 5-1 by Kuulkers et al.
(1994), and is thought to relate to 
X-ray dipping. As the aim of the present work is to understand
the normal Z-track, a selection was made to remove about 800 s of data showing
this behaviour. 

\subsection{Spectral analysis}

The Z-track shown in Fig. 2 is derived from lightcurves with 64 s binning.
Increasing the bin size from the intrinsic 16 s binning of PCA Standard 2 data
reduces the width of the Z-track at all points. More important, our previous
work (e.g. Barnard et al. 2003) has shown that data should be selected 
on a line through the centre of the Z-track as we are interested in the changes taking
place along the track and not perpendicular to it. 
We select spectral data at positions about equally spaced along
the Z-track by defining boxes in the hardness-intensity plane and selecting data
within each box. A good time interval (GTI) filter was produced for each selection
consisting of the time intervals when data were inside each box
and used for extraction of lightcurves and spectra.
A hardness-intensity plot was then made for the selected data
and overlayed on the full Z-track to see how good the selection was, checking
that points lie on a smooth curve not deviating appreciably perpendicular
to the Z-track. 

\begin{figure*}                                                          
\includegraphics[width=60mm,height=86mm,angle=270.]{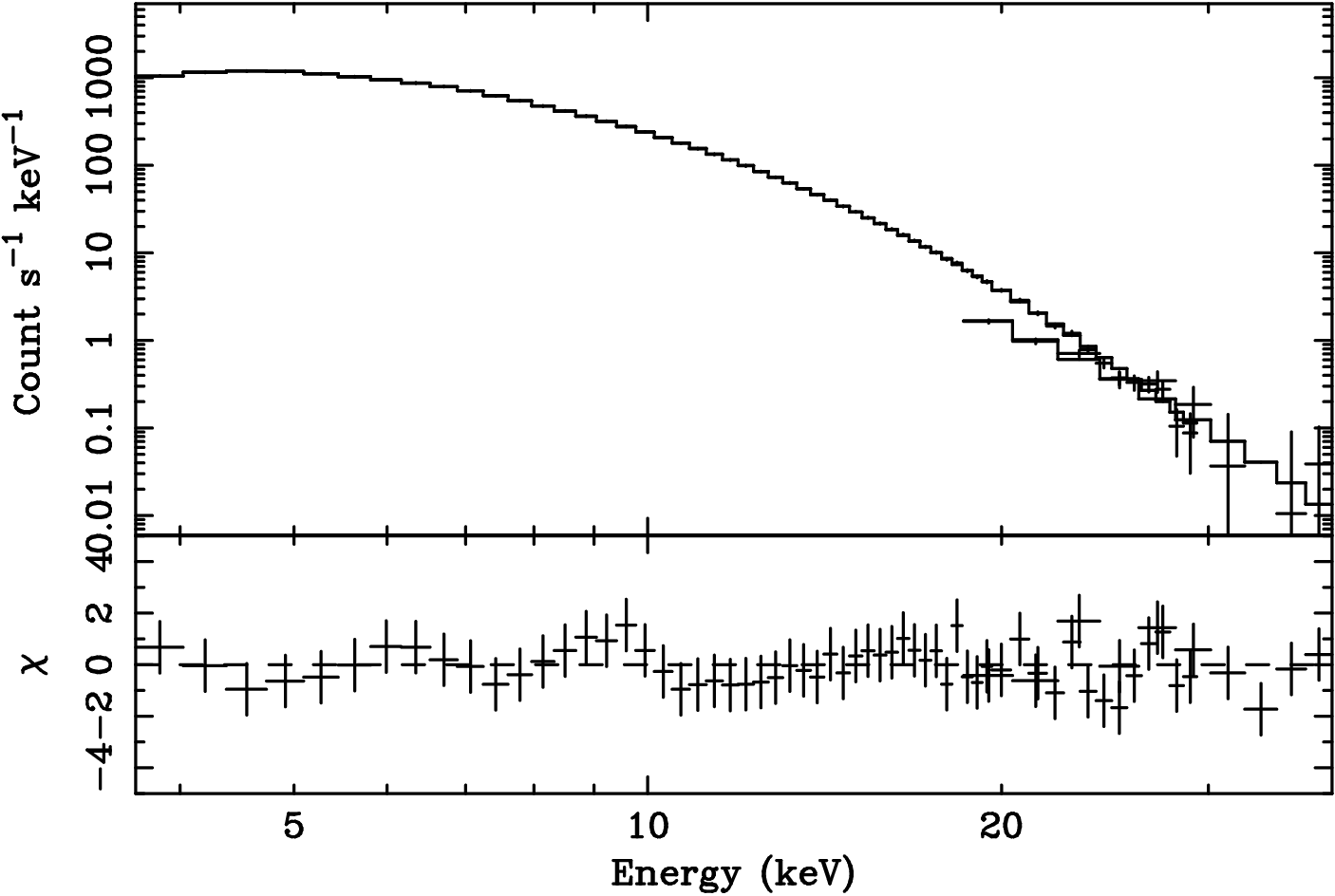}    
\includegraphics[width=60mm,height=86mm,angle=270.]{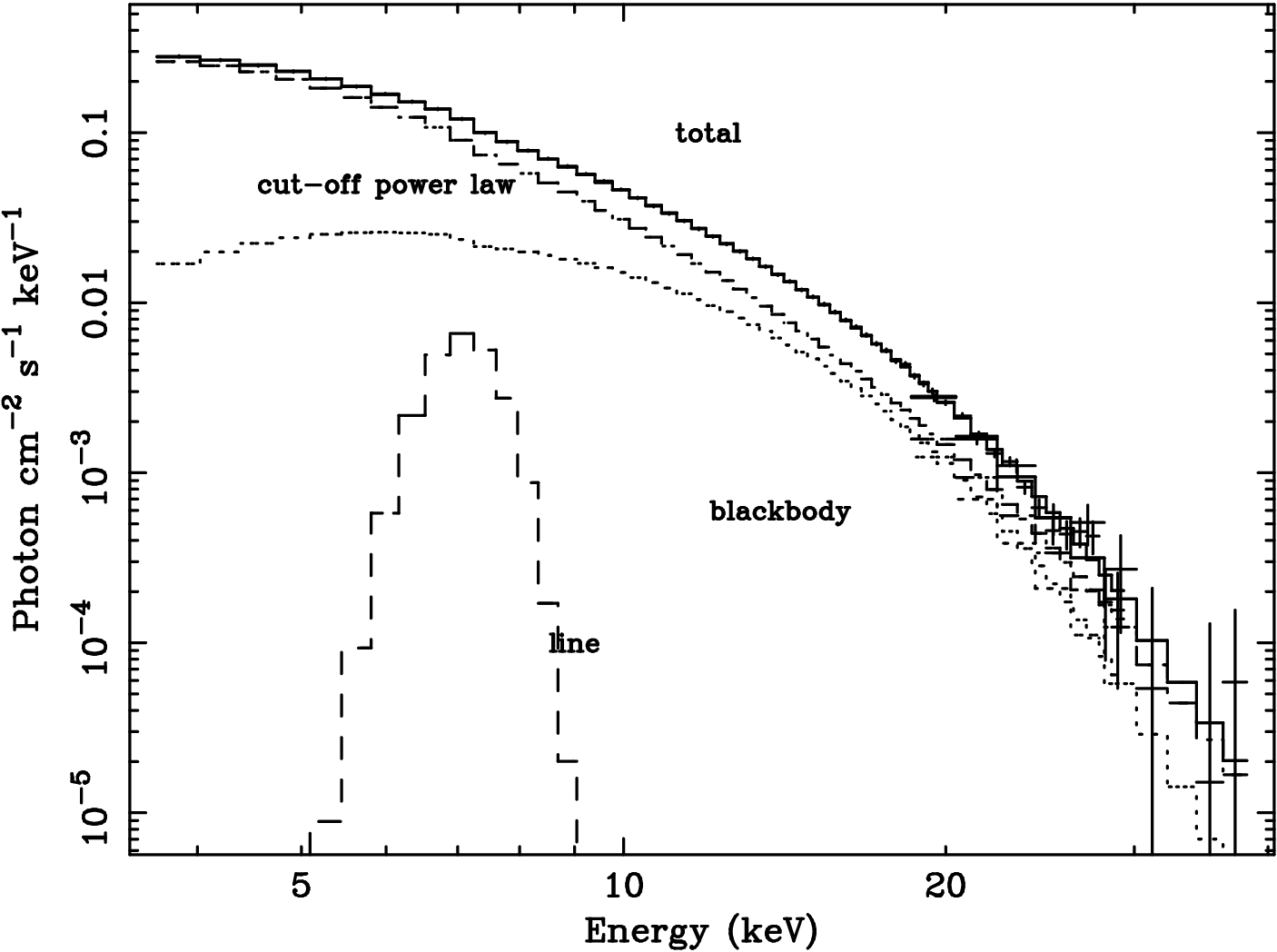}    
\includegraphics[width=60mm,height=86mm,angle=270.]{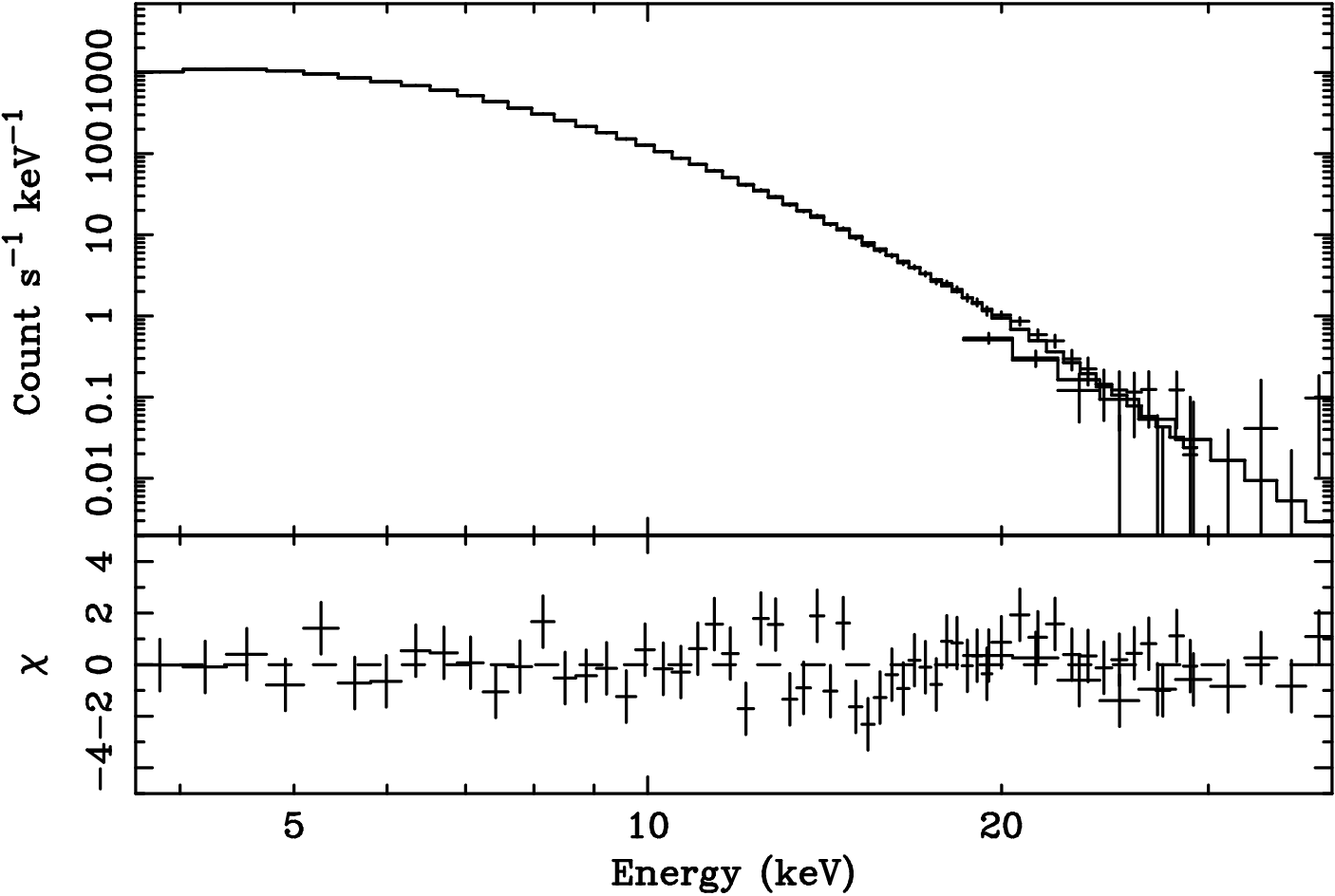}    
\includegraphics[width=60mm,height=86mm,angle=270.]{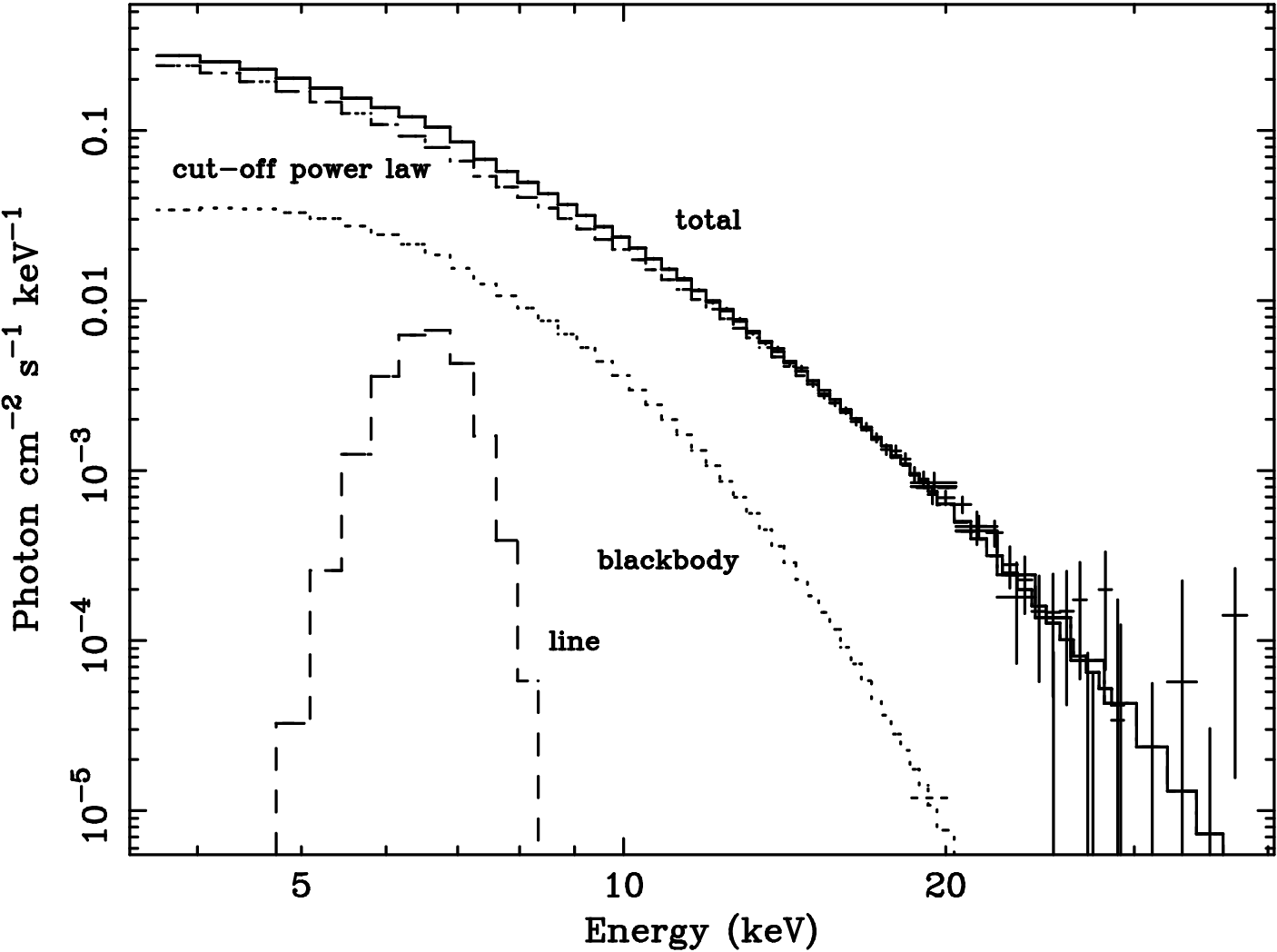}    
\includegraphics[width=60mm,height=86mm,angle=270.]{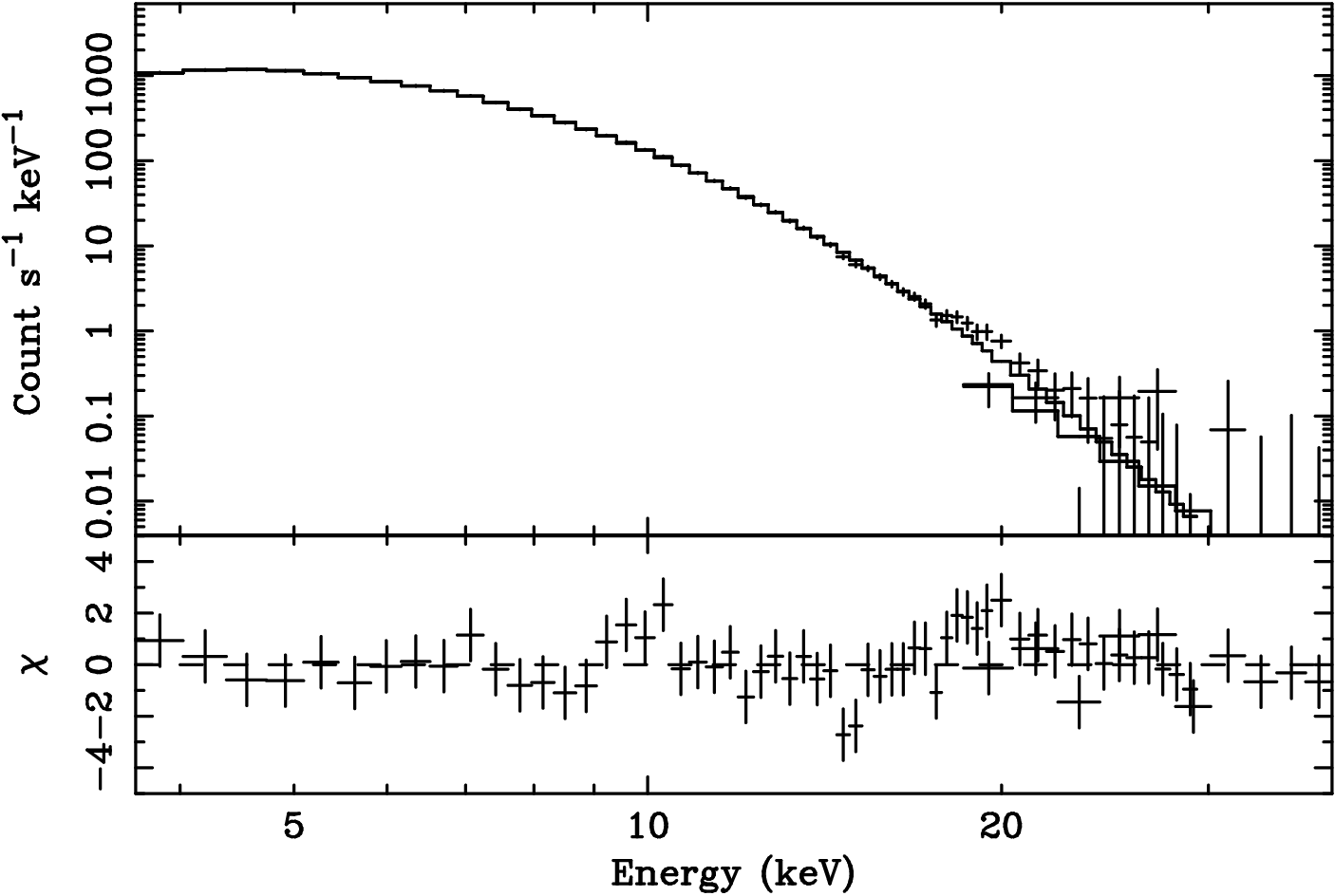}    
\hskip 8 mm
\includegraphics[width=60mm,height=86mm,angle=270.]{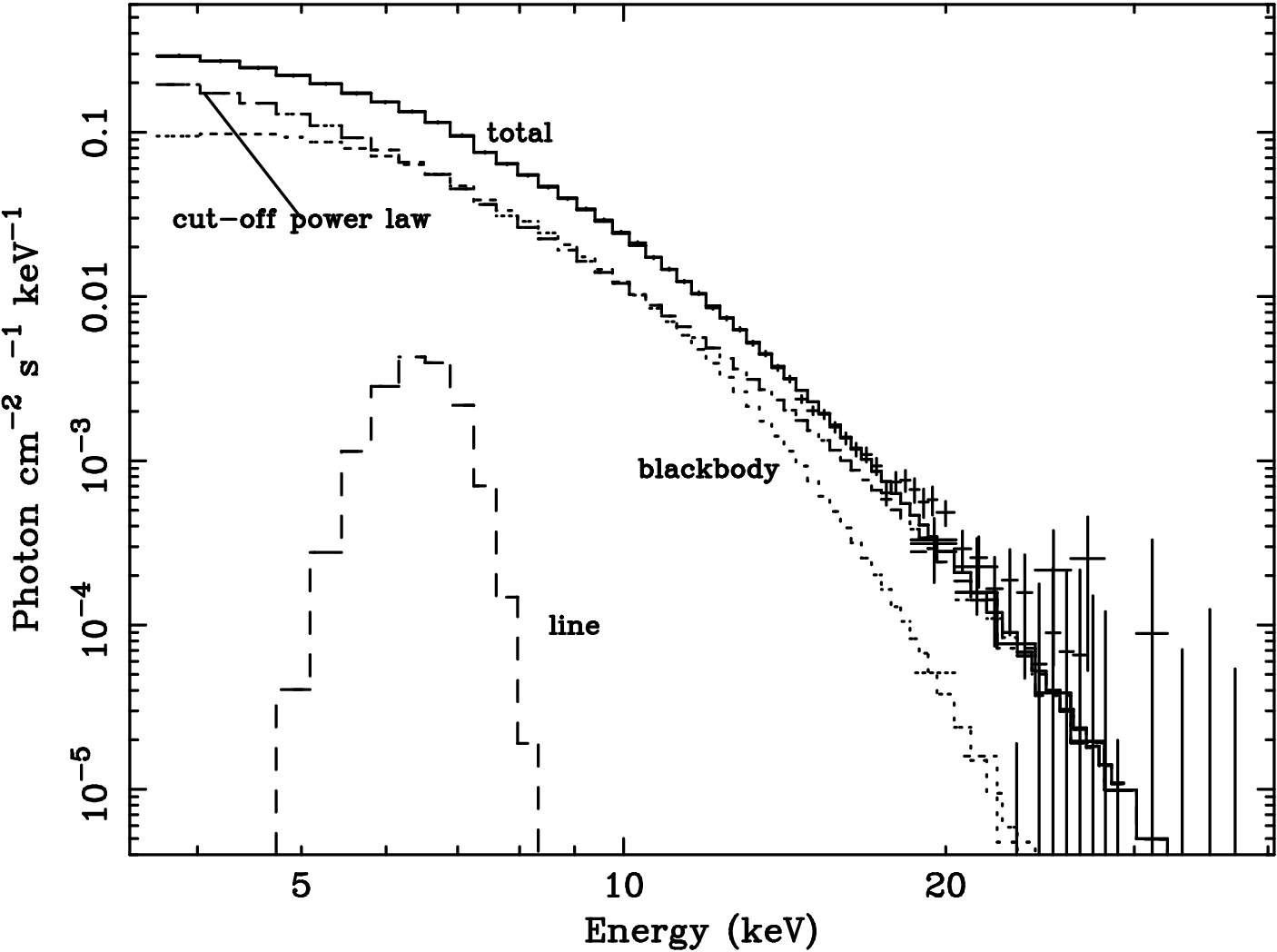}    
\caption{Spectra for the HB (top; spectrum 2), NB (centre; spectrum 6) and FB (bottom; spectrum 9).
Both the PCA and HEXTE data are shown, in each case with the folded data (left) with residuals and 
unfolded data (right). Over large parts of each spectrum the error bars are too small to be visible. 
Model components are labelled in the unfolded spectra.}
\label{}
\end{figure*}

A total of 10 spectra were extracted with three spectra on the HB (one at the hard apex),
four on the NB including one at the soft apex, plus three on the FB. The ranges of intensity
were generally 100 count s$^{-1}$ wide, e.g. from 6050 -- 6150 count s$^{-1}$ with a range of
hardness ratio in this case from 0.425 -- 0.435. For each selection, PCA spectra were
extracted and the corresponding background spectra generated. Deadtime correction was
made to source + background (S + B) spectra and to background (B) spectra using a local
facility {\sc pcadead}. Pulse pileup correction was 
found to have negligible effect and so was not made. A systematic error of 1\% 
was applied to each channel as is standard practice in analysing PCA data;
it was not necessary to regroup data to a minimum count per
bin for use of the $\chi^2$ statistic as the count in raw channels was already 
high (e.g. typically 200). 

\begin{table*}
\caption{Spectral fitting results: column density, blackbody temperature $kT_{\rm BB}$, normalization 
and blackbody radius $R_{\rm BB}$; power law cut-off energy $E_{\rm CO}$ and normalization; 
line energy $E_{\rm l}$ and equivalent width; goodness of fit. 90\% confidence errors are shown.}
\begin{center}
\begin{minipage}{160mm}
\begin{tabular}{lrrrrlrrrrr}
\hline\noalign{\smallskip}
$\;\;$spectrum&$N_{\rm H}$&$kT$&$norm_{\rm BB}$&$R_{\rm BB}$&$E_{\rm CO}$&$norm_{\rm CPL}$&$E_{\rm l}$&EW&$\chi^2$/d.o.f.\\
&&keV&&km&keV&&keV&eV\\
\noalign{\smallskip\hrule\smallskip}

Horizontal Branch\\
1&9.2$\pm$0.8 &2.44$\pm$0.11 &3.6$\pm$0.4  &3.1$\pm$0.3  &5.8$\pm$0.3 &9.13$\pm$0.9   &6.9$\pm$0.4&81 &27/59\\
2&10.2$\pm$0.8 &2.48$\pm$0.09 &4.3$\pm$0.6  &3.3$\pm$0.3  &5.2$\pm$0.4 &11.7$\pm$0.8   &7.1$\pm$0.4&75 &41/59\\ 
3&10.6$\pm$0.8 &2.10$\pm$0.21 &2.2$\pm$0.5  &3.3$\pm$0.8  &5.4$\pm$0.2 &13.9$\pm$1.3   &7.1$\pm$0.4&68 &39/59\\
\hline\noalign{\smallskip}
Normal Branch\\
4&9.5$\pm$0.8 &1.70$\pm$0.18  &2.1$\pm$0.6  &4.9$\pm$0.6  &5.2$\pm$0.2 &12.8$\pm$1.4  &6.7$\pm$0.4&79  &29/59\\
5&8.8$\pm$0.9 &1.32$\pm$0.09  &2.46$\pm$1.0  &8.6$\pm$1.8  &4.9$\pm$0.2 &11.4$\pm$1.5  &6.6$\pm$0.3&80  &27/59\\
6&8.4$\pm$0.5 &1.32$\pm$0.05 &2.8$\pm$0.8   &9.4$\pm$1.5  &4.3$\pm$0.1 &10.9$\pm$1.1  &6.6$\pm$0.1&84  &63/59\\
7&8.0         &1.29$\pm$0.03 &3.5$\pm$0.6   &10.8$\pm$1.1 &3.7$\pm$0.1 &10.8$\pm$1.2  &6.5$\pm$0.1&91  &92/60\\
\hline\noalign{\smallskip}
Flaring Branch\\
8&8.0         &1.40$\pm$0.05 &6.0$\pm$0.6  &12.2$\pm$1.1  &3.4$\pm$0.2 &11.0$\pm$1.1    &6.4$\pm$0.3&64  &65/60\\
9&8.0         &1.39$\pm$0.04 &8.0$\pm$0.7  &14.3$\pm$0.8  &3.6$\pm$0.2 &10.2$\pm$1.1    &6.5$\pm$0.4&45  &70/60\\
10&8.0        &1.42$\pm$0.03 &11.1$\pm$0.5 &16.1$\pm$0.8  &3.3$\pm$0.2 &10.3$\pm$1.3    &6.4$\pm$0.8&14  &63/60\\
\noalign{\smallskip}\hline
\end{tabular}\\
Column densities are in units of 10$^{22}$ atom cm$^{-2}$; the normalization of the blackbody is in units 
of $\rm {10^{37}}$ erg s$^{-1}$ for a distance of 10 kpc, the normalization of the cut-off power law is in units of
photon cm$^{-2}$ s$^{-1}$ keV$^{-1}$ at 1 keV and the line normalization has units of photon cm$^{-2}$ s$^{-1}$.
\end{minipage}
\end{center}
\end{table*}

HEXTE spectra were also extracted from Cluster 1 for detectors 0, 1 and 3, detector
number 2 having failed in March 1996. Data were selected using the
GTI filter files generated from PCA data using the {\it RXTE} {\sc ftool} {\sc hxtlcurv}
which also deadtime corrected the output files.
This produced a source + background and a background spectrum from each raw data file, which were then
added to form a single S + B spectrum and a single B spectrum. The auxiliary instrument response
file (arf) of May 2000 was used, together with the response matrix file (rmf) of March 1997. The rmf file
was rebinned to match the actual numbers of channels in the HEXTE spectra using the {\sc ftools}
{\sc rddescr} and {\sc rbnrmf}. The energy ranges used in the fitting were set by examining
the energies at which the spectral flux density of the background became equal to that of
the source + background. Typically, the ranges used were 3.5 -- 30 keV in the PCA and 18 -- 40 keV in HEXTE. 

The aim of this work was to determine whether application of a particular emission model 
(the Birmingham model) would provide a good fit to spectra and 
suggest the nature of the changes taking place,
and so an explanation of the Z-track. Thus it was not the intention to fit a variety
of two-component models as it is widely appreciated
that various models can give acceptable fits, and it is not sensible to try to choose between
physical models on the basis of rather small differences in $\chi^2$. In particular, as
discussed in Section 1, the evidence against the Eastern model is now strong and so
in the present work
the Birmingham model was applied but with extensive testing of various aspects of the fitting.

Firstly, it was clear from examining the residuals when a continuum only model was applied
that a broad iron line was present, and the width of the peak in the residuals suggested
a half-width of the line $\sigma $ of $\sim$0.5 keV. Fitting with and without a line
made the significance of the line clear: without a line, spectrum 2 gave a $\chi^2$/d.o.f.
of 66/61 improving to 46/60 for a line at energy 6.5 keV. In the final fitting, the line energy
was free and a systematic variation around the Z-track was seen (Sect. 3.2.3), and for
spectrum 2, freeing the energy further improved $\chi^2$/d.o.f. to 41/59.
With the inclusion of a
Gaussian line in the spectral model, good values of $\chi^2$/d.o.f were obtained at all
positions on the Z-track for the model {\sc bb + cpl + gau}; i.e. a simple blackbody
from the neutron star, Comptonized emission specified by a cut-off power law as 
discussed in Sect. 1,  plus the Gaussian line (the emission site of the line is discussed
in Sect. 3.2.3). For the 
blackbody component, the fitting providing well-determined values for
the blackbody temperature $kT_{\rm BB}$ and normalization. From the normalization,  defined
as the luminosity adjusted to a source distance of 10 kpc, the blackbody radius $R_{\rm BB}$ 
was derived via $L_{\rm BB}$ = 4$\pi R_{\rm BB}^2 \sigma T_{\rm BB}^4$ 
where $L_{\rm BB}$ is the luminosity of the blackbody component and $\sigma$ is Stefan's constant.
A clear pattern of behaviour of $kT_{\rm BB}$ and $R_{\rm BB}$ along the Z-track emerged, and 
there was also a clear pattern of behaviour of the luminosities $L_{\rm BB}$
and $L_{\rm CPL}$. Thus the spectral fitting results were robust. 

With high luminosity sources such as GX\th 340+0, the Comptonized
emission does not extend to very high energies as the cut-off energy $E_{\rm CO}$
is relatively low, i.e. a few keV (e.g. present work; Barnard et al. 2003)
when compared with typical X-ray burst sources
such as XB\th 1916-053 with much smaller luminosities. In the case of such
sources, broadband spectra especially from the {\it BeppoSAX} satellite extending from
0.1 -- 100 keV or more can be used to provide $E_{\rm CO}$ and the value of the power
law photon index $\Gamma$. In XB\th 1916-053,  $E_{\rm CO}$ was found to be 80$\pm$10 keV
(Church et al. 1998b). In the present data it was found that $E_{\rm CO}$ was well-determined
in the spectral fitting, varying between 3 and 6 keV depending on position in the Z-track.
However, the low value of cut-off energy restricts the range of energy available for
determination of the power law index and the procedure was adopted of fixing $\Gamma$ at
1.7. This value is physically reasonable as it corresponds to a low Comptonizing region
temperature and high optical depth (Shapiro et al. 1976).
After fitting to obtain all other parameters, the index was freed, and it was found
that it remained close to 1.7. Extensive testing revealed that the pattern of behaviour
shown in Figs. 4 -- 6 was always the same and did not depend on the value of the power law index.
Similarly, we adopted the procedure previously
used (e.g. Barnard et al. 2003) when an emission line is broad and 
fixed the Gaussian half-width $\sigma$ at 0.5 keV 
to prevent the line absorbing neighbouring continuum. 
Good fits were obtained to all spectra; $\chi^2$/d.o.f. appears 
marginally high for spectrum 7, but it is difficult to be sure of the precise level of
systematic error to use for a bright source in which Poisson errors are negligible. 
The systematic error added in general to data from a given observatory is based on calibrations
and an instrument response that is improved during the mission. However, calibrations
are not carried out using the brightest sources, and so the 1\% used here may be too low,
as we previously found in the case of Sco\th X-1 (Barnard et al. 2003), 
and if systematic errors of 1.5\% are used, $\chi^2$/d.o.f. becomes 57/60 for this spectrum.

The low energy limit of the PCA instrument (normally set in spectral fitting at
$\sim$3 keV) mitigates against determination of the column density $N_{\rm H}$, especially 
if the Galactic column density is low. It is often the case in fitting {\it RXTE} spectra
that the value of $N_{\rm H}$ has to be fixed at the known Galactic value.
In the present data, acceptable fits were only obtained with values of
$N_{\rm H}$ $\sim 8\times 10^{22}$ atom cm$^{-2}$, several times higher than the Galactic
value of $\sim 2.2\times 10^{22}$ atom cm$^{-2}$ (Dickey \& Lockman 1990), suggesting 
absorption intrinsic to the source. (When we fitted the Eastern model to a number of
spectra, an even higher column density of $\sim 12\times 10^{22}$ atom cm$^{-2}$ was needed.)
With the Birmingham model, fitting with $N_{\rm H}$ fixed at $8\times 10^{22}$ atom cm$^{-2}$
produced acceptable fits for all spectra. However, careful investigation revealed a
systematic change of $N_{\rm H}$ along the Z-track, with the column density increasing
along the normal branch towards the hard apex and then decreasing on the horizontal branch.
Thus  $N_{\rm H}$ was allowed to be free except for the spectra on the flaring branch 
(spectra 7 -- 10) where somewhat reduced count statistics limited the ability to determine $N_{\rm H}$.
The effect in these spectra of fixing $N_{\rm H}$ on the other parameters was small,
varying from 1\% in one spectrum to less than 15\% in another.
The final fitting results are shown in Table 1.
It can be seen from the 90\% confidence limits for $N_{\rm H}$ that
the change in column density around the Z-track is significant.

\subsubsection{The neutron star blackbody}

Results for the blackbody parameters $kT_{\rm BB}$ and $R_{\rm BB}$ as a function of position
on the Z-track are shown in Fig. 4. It can be seen from the small size of the 90\% confidence error 
bars that these parameters are well constrained, as is their evolution around the Z-track.
The luminosities of the two continuum components
are given in Fig. 5, in all cases as a function of the total luminosity
\begin{figure}
\begin{center}
\includegraphics[width=86mm,height=86mm,angle=270]{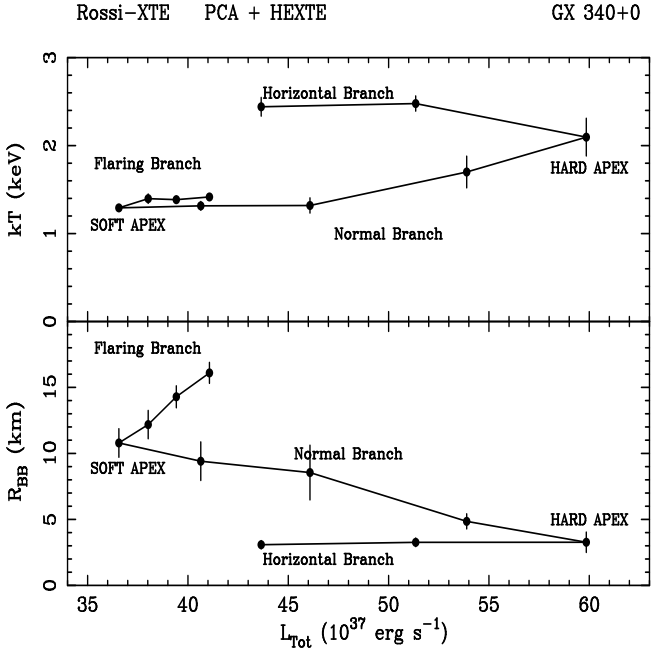}                   
\caption{Variation of neutron star blackbody temperature and radius along the Z-track
showing the monotonic changes of each on both the Normal and Horizontal Branch,
i.e. with no sudden change at the Hard Apex, unlike the non-thermal emission (see Fig. 6).}
\end{center}
\end{figure}
\begin{figure}
\begin{center}
  \includegraphics[width=86mm,height=86mm,angle=270]{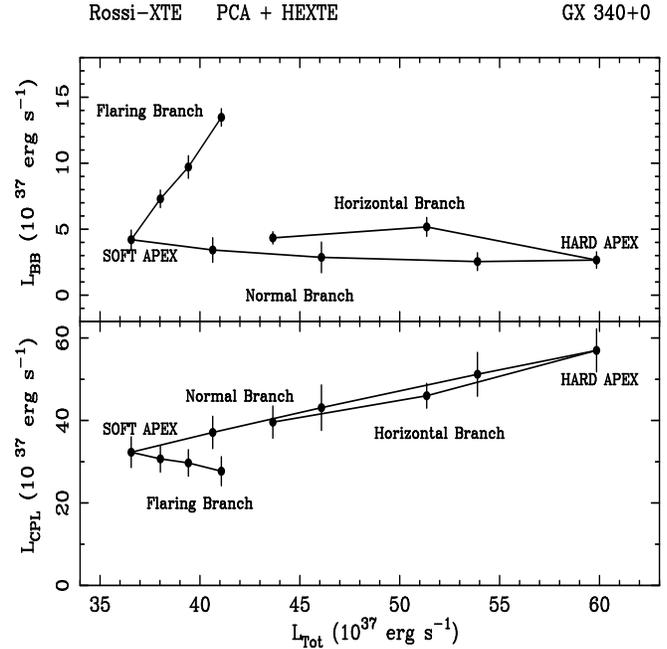}                
\caption{Variation of the luminosities of the two continuum components along the Z-track
showing the marked increase of the Comptonized emission on the Normal Branch and a
similar decrease on the Horizontal Branch.}
\end{center}
\end{figure}
$L_{\rm Tot}$ in the band 1 -- 30 keV from the unabsorbed flux of the best-fitting 
spectral model for each spectrum. A source distance of 10 kpc was assumed 
based on the considerations of Christian \& Swank (1997) who point out that as a
bright Galactic source with coordinates (339$\degmark$.6 -0$\degmark$.1) 
it is likely to be at $\sim$10 kpc
and who found an upper limit of 11 kpc for the distance. The distance is unlikely
to be less than 5 kpc as this would be inconsistent with the high column density
of the source even allowing for a large amount of intrinsic absorption, and 
because an optical counterpart would have been detectable. Thus for a distance
uncertainty of -50\% +10\%, absolute luminosities 
will be known to better than a factor of two, while the pattern of the parameter
variations is Figs. 4 -- 6 will not be affected at all by systematic uncertainty 
in luminosity. From Fig. 4, it
can be seen that the blackbody parameters have a
systematic evolution along the Z-track. Firstly, the blackbody temperature is smallest
at the soft apex at $\sim$1.25 keV. From this point, the temperature increases as we
proceed towards the hard apex, and continues to increase on the horizontal 
branch. The most striking result is the value of the blackbody radius $R_{\rm BB}$ 
which is $\sim$11 km at the soft apex (10.8$\pm$1.1 km). In addition to the 10\%
error from spectral fitting there will be a possible systematic error due to the
uncertainty in source distance. $R_{\rm BB}$ decreases continually along the
normal branch and the horizontal branch. On the flaring branch, there is a small but significant
increase of temperature, but a marked increase of blackbody radius to 16 km.
Moreover, in the non-flaring source, $R_{\rm BB}$ is a maximum 
at the soft apex. The most obvious explanation is that this is the natural
limit when the whole surface of the neutron star is emitting, while away from this point, the whole
neutron star is not emitting.

On the basis of these results we propose that the soft apex is a quiescent state of the 
source with the neutron star emitting with its lowest temperature and from the whole
of its surface. We next discuss the changes taking place on moving away from the quiescent state of the source
and for this, we need to consider the luminosities of the continuum components.
Fig. 5 shows the evolution along the Z-track of the luminosities of the 
individual blackbody and Comptonized continuum components
as a function of the $L_{\rm Tot}$. The major component
is the Comptonized emission which varies between 
30 -- 55$\times 10^{37}$ erg s$^{-1}$ on the normal branch while $L_{\rm BB}$ is substantially less.

Firstly, consider motion away from the quiescent state towards the
hard apex. The most striking result is the strong increase in the luminosity of the Comptonized 
emission $L_{\rm CPL}$, followed by a decrease on the horizontal branch
by about the same amount essentially back along the same path.
\begin{figure}
\begin{center}
\includegraphics[width=86mm,height=86mm,angle=270]{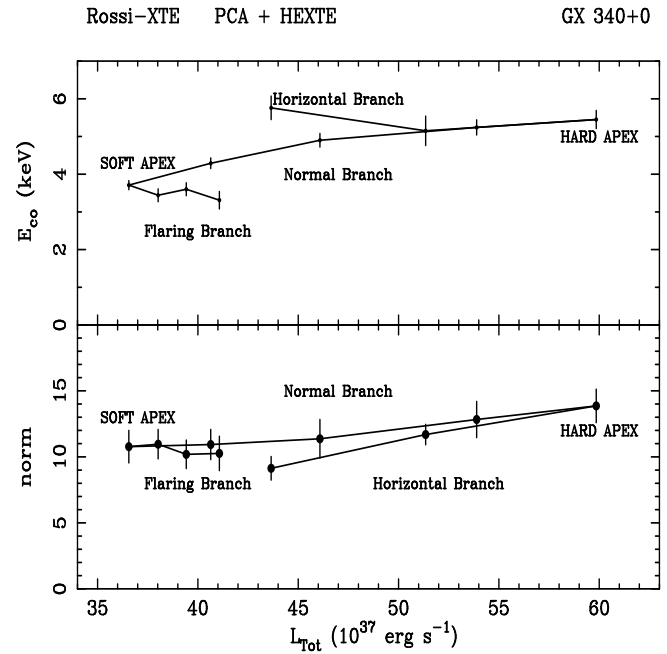}                        
\caption{Variation of the normalization and cut-off energy of the Comptonized 
emission along the Z-track.}
\end{center}
\end{figure}
Application of the spectral model used in the present work has
revealed this for the first time. The Comptonized emission
arises in the ADC located above the accretion disc 
and it is expected that the ADC emission will 
be related to the properties, particularly the density of the accretion disc,
depending on the mass accretion rate $\dot M$. We suggest that 
it is {\it highly unlikely} that this strong increase of luminosity of the Comptonized emission 
could take place without an increase of $\dot M$, and so propose that the evolution 
of $L_{\rm CPL}$ is caused by an increase of $\dot M$ on the normal branch. 
The results (Fig. 5) imply that it is followed by a decrease of $\dot M$ back to 
about the initial value on the horizontal branch. We point out here that under the conditions
of high radiation pressure that exist at the hard apex and HB (below), the measured value of
$L_{\rm Tot}$ may not give the true $\dot M$ as appreciable mass accretion can be
diverted into the jets such that the mass accretion rate may not actually decrease when
the source gets to the hard apex, but that this may happen at some later time.
This is discussed in Sect. 4.2.

This does not support the previously held view that $\dot M$ changes monotonically 
along the Z-track (increasing in the direction HB -- NB -- FB), since we propose that 
the mass accretion rate increases from the soft apex to the hard apex.
If this is so, then it suggests that the evidence from 
previous work of an increase in UV or optical emission around the Z-track (Sect. 1.3) 
is not conclusive, or if an increase were confirmed, that this does not imply an increase 
of mass accretion rate. The blackbody luminosity behaves quite differently from that
of the Comptonized emission not reversing its trend at the hard apex
but continuing to increase on the horizontal branch reflecting the smooth variation 
of $kT_{\rm BB}$ which increases on both the normal and horizontal branches, and 
of $R_{\rm BB}$ which decreases on both branches (Fig. 4). This suggests 
that the same process is affecting the blackbody emission on the two branches.

Fig. 4 shows that the blackbody temperature $kT_{\rm BB}$ increases from $\sim$1.3 keV
in the quiescent state of the source at the soft apex to 2.0 keV at the hard apex, and
2.4 keV on the horizontal branch, and we propose that this increase is due to an
increasing mass accretion rate. Most of the change takes place between the upper half
of the normal branch and halfway along the horizontal branch. Beyond this point the
temperature appears to have reached a stable value. Similarly, the blackbody radius
decreases from $\sim$11 km at the soft apex to $\sim$3 km on the horizontal branch.
The change in $kT_{\rm BB}$ by about a factor of two will mean that the radiation pressure
of the neutron star emission varying as $T^4$ will increase by nearly a factor of
10. At the same time the emitting area shrinks from the whole neutron star down to a
narrow equatorial belt. The half-height $h$ of this belt is calculated from $R_{\rm BB}$
since the area $A$ = $4 \pi R_{\rm BB}^2$ = $4 \pi h R$ (the surface area of a
sphere intersected by two parallel planes 2$h$ apart, where $R$ is the radius
of the neutron star assumed to be 10 km). On the horizontal branch, $h$ is close to 1.0 km.
The inner disc in bright sources is radiatively-supported, i.e. not thin, but extends
vertically tens of kilometres above the orbital plane (Frank et al. 2002).
We propose that the strong radiation pressure of the neutron star removes the outer layers of the inner 
disc (acting close to vertically upwards) while having little effect on the disc in 
the orbital plane where it acts horizontally into the disc.
In Sect. 4.1 we show that the decrease in blackbody area is consistent with  
disruption of the inner disc. In addition, the observed increase of column density
provides direct evidence for disruption of the inner disc, or at least for release
of material somewhere within the source. Detailed modelling is desirable to demonstrate
the disruption of the inner disc proposed.

Moreover, we can see the strength of the increased radiation pressure
by comparison with the Eddington limit.
For the source at the horizontal branch end of the Z-track, $L_{\rm BB}$ 
is $4.3\times 10^{37}$ erg s$^{-1}$, i.e. 0.25 $L_{\rm Edd}$ while the total luminosity
is $4.4\times 10^{38}$ erg s$^{-1}$. Using the blackbody radius of 3.08 km we find
that the flux emitted by the surface $f$ is $3.6\times 10^{25}$ erg cm$^{-2}$ s$^{-1}$.
On the surface of the star at $r$ = $R$, the critical Eddington flux
$f_{\rm Edd}$ is $L_{\rm Edd}/4\, \pi\, R^2$, i.e. $1.4\times 10^{25}$ erg cm$^{-2}$ s$^{-1}$
(for a neutron star radius of 10 km). Thus 
at this position on the Z-track, the flux is super-Eddington and when we are close to the
narrow equatorial strip we may expect strong effects due to the high radiation pressure.
By contrast, at the 
soft apex, the blackbody luminosity is similar at $4.2\times 10^{37}$ erg s$^{-1}$ 
to that at the end of the Z-track, but the
flux emitted is fifteen times reduced at $0.24\times 10^{25}$ erg cm$^{-2}$ s$^{-1}$,
substantially sub-Eddington.
The variation of
$f/f_{\rm Edd}$ along the Z-track is given in Fig. 7 showing that the ratio increases
to more than unity on the normal branch reaching a maximum on the horizontal branch.

\begin{figure}
\begin{center}
\includegraphics[width=86mm,height=86mm,angle=270]{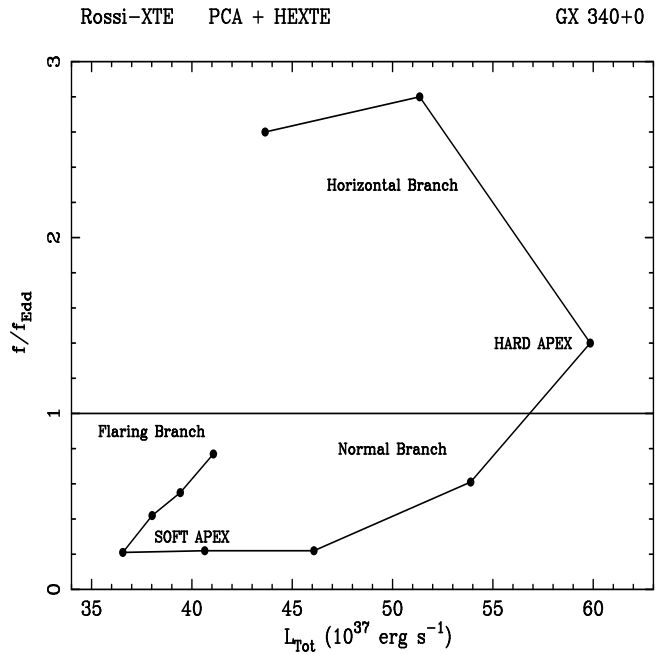}                   
\caption{The neutron star emitted power per unit area relative to the Eddington flux showing that the
effects of radiation pressure are strong on the upper normal and horizontal branches.}
\end{center}
\end{figure}

Moving from the 
soft apex onto the flaring branch, the most obvious feature is the approximate constancy 
of the Comptonized emission luminosity. This is strongly suggestive that the mass
accretion rate does not change on this branch, since any significant change would
surely result in $L_{\rm CPL}$ changing. However, there are strong changes in the
neutron star blackbody with the luminosity increasing by a factor of three.
While $kT_{\rm BB}$ displays only a small increase of less than 10\%, there is a 
large increase in emitting area (blackbody radius). The nature of flaring in such sources has been 
controversial. The view of a monotonically increasing $\dot M$ around the Z-track
implies that an increase of mass accretion rate causes flaring. Alternatively, 
the flares may be thermonuclear and the present results with implied constant
$\dot M$ suggest this quite strongly. Thus lower luminosity LMXB often
exhibit unstable burning as X-ray bursting while brighter sources exhibit flaring.
The implied constancy of $\dot M$ on the flaring 
branch again disagrees with the often held view that $\dot M$ increases
monotonically on the Z-track in the direction HB--NB--FB. 

On the flaring branch, the blackbody radius increases to 16 km so that if 
the whole neutron star is emitting at the soft apex, then the emission
may spread beyond the neutron star in flaring, as discussed in Sect. 4.3.
The radiation pressure also increases. From the soft apex to the spectrum near the peak 
of flaring (spectrum 10) $L_{\rm BB}$ increases by a factor of three (due to
$R_{\rm BB}$ increasing). Allowing for the area increase, the emitted flux increases by 50\% while the Eddington flux 
$L_{\rm Edd}/4\, \pi\, R^2$ decreases to $5.4\times 10^{24}$ erg cm$^{-2}$ s$^{-1}$ for 
spectrum 10, so that the $f/f_{\rm Edd}$ increases from 0.24 at the soft apex to 
0.77 in spectrum 10, and is presumably close to 1.0 at the actual flare peak, for
which there was no spectrum (see Fig. 7). 

\begin{figure}                                                           
\includegraphics[width=86mm,height=86mm,angle=270]{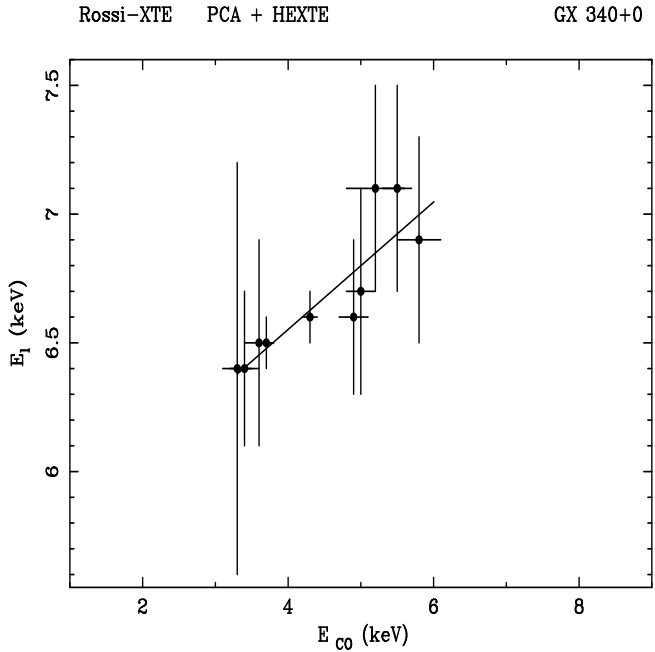}
\caption{Correlation of Fe line energy with the cut-off energy of the Comptonized emission.}
\label{}
\end{figure}

\subsubsection{The Comptonized emission}                                              
The variation of the cut-off energy and the normalization of the Comptonized emission
are shown in Fig. 6. Both of these were well-determined in the fitting. The normalization, of 
course, mirrors the behaviour of $L_{\rm CPL}$. The cut-off
energy $E_{\rm CO}$ is correlated with the blackbody
temperature, suggesting a link between the neutron star emission
and the properties of the ADC. This is certainly true on the normal and horizontal 
branches. The cut-off energy reflects the electron temperature $kT_{\rm e}$ of the ADC
plasma, such that for high optical depth to electron scattering $E_{\rm CO}$ = $3kT_{\rm e}$
(e.g. Petrucci et al. 2001).
The observed correlation suggests that the ADC electron temperature is responding to changes 
in the neutron star temperature. 

Our previous work is consistent with a single ADC electron temperature: if the ADC temperature
varied substantially as the disc temperature does, say by a factor of ten, we would expect
the knee at the Comptonization cut-off to be spread over a factor of ten in energy, e.g. from
2.5 to 25 keV which would be very obvious. We have never detected any departure from a
single cut-off energy in any of our work on LMXBs (e.g. Church et al. 1997).
Moreover, there is other evidence that the ADC is isothermal.
In lower luminosity sources, the cut-off energy is generally high (e.g. di Salvo \& Stella 
2002) implying a high electron temperature of 25 keV or more, 
although this is not understood since theoretically $kT_{\rm e}$ should
be limited to the inverse Compton temperature, i.e. less than $\sim$3 keV. Thus there must
be a presently unknown process heating the corona. In brighter sources 
($L$ $>$ $10^{37}$ erg s$^{-1}$), the cut-off
energy falls to a few keV, and the neutron star and ADC temperatures 
are equal (Ba\l uci\'nska-Church \& Church 2005) indicating thermal
equilibrium between the neutron star and ADC which could be due to the higher plasma density 
in the ADC giving high thermal conductivity. This also suggests that there is thermal equilibrium within the ADC
resulting in the ADC being isothermal. Thus, the use of a simple cut-off power law
within the Birmingham model is justified on the basis of an isothermal ADC.

In the present data on GX\th 340+0, the average ratio of $E_{\rm CO}$/$kT_{\rm BB}$
from the spectral fitting results is 2.49$\pm$0.46 at 67\% confidence. The constancy
of the ratio along the Z-track
is striking and the closeness of the average to 3.0 implies that the ADC 
temperature is equal to the blackbody temperature if the plasma
has high optical depth. 

\subsubsection{The line emission}

Table 1 shows the dependence of the observed iron line energy on Z-track position with the lowest
energy at the soft apex, systematically increasing along the NB to the hard apex, 
remaining high on the HB. There is a good correlation of the line energy with $E_{\rm CO}$, as shown 
in Fig. 8, since a linear fit to these data taking into account the 90\% confidence errors in both 
$y$ and $x$ gives a slope of 0.25 $\pm$0.04 which is inconsistent with zero slope, and the correlation 
coefficient is 0.985. The cut-off energy is a measure of the electron temperature in the ADC, and 
Fig. 8 provides clear evidence for the origin of the line in the ADC, important because 
the site of line emission in LMXBs has been controversial. The line energy at the soft apex of $\sim$6.5 keV 
suggests a low ionization state while at the hard apex for the energy of $\sim$7 keV
the state is high (Makishima 1986). The ionization state will increase if the ionization
parameter $\xi$ = $L/n\,r^2$ increases. At the soft apex the whole neutron star emits isotropically,
but at the hard apex, the radiation from an equatorial strip becomes concentrated towards the
disc and ADC so increasing $\xi$ substantially, and this would explain the increased line energy 
at the hard apex.

There is little evidence for the line normalization changing systematically, and so
the equivalent width (EW) appears approximately constant. However, there is an apparent decrease of EW 
in flaring. The normal calculation of EW compares the line flux with the total 
continuum flux, and the continuum emission of the ADC where the line appears to be generated
has constant luminosity in flaring. If the EW is calculated on the basis of this ADC continuum component
and not the total, there is no decrease in EW (in spectra 8 and 9) and only a possible
decrease for spectrum 10.

\section{Discussion}

We have investigated spectral changes along the Z-track based on the use of a particular emission 
model. In the following we examine the implications of the results on the assumption that this
model is correct. As discussed in Sect. 1, we do not know of any major inaccuracy in use
of the spectral form {\sc bb + cpl} within the model, and it is difficult to quantify any small level of inaccuracy.
However, such inaccuracy will not change the {\it pattern} of parameter variation around the
Z-track and is not expected to lead to large errors in, for example, the blackbody radius
at the soft apex.
The results provide a plausible physical picture of the Z-track, of the observation of jet radio
emission on the horizontal branch and of the nature of flaring, which is quite different from
previous pictures. In the model presented here, the constant luminosity of the ADC emission 
on the FB implies $\dot M$ is constant and so flaring must be thermonuclear. The soft apex is the quiescent 
state of the source, with $\dot M$ increasing along the normal branch causing an increase of the 
ADC luminosity. The neutron star blackbody temperature then rises causing increased radiation 
pressure which has a strong effect on the inner disc, blowing away the outer layers (farthest
from the orbital plane). It is
not appropriate to discuss our model in relation to QPO in detail here, however, we note that
the blowing away is not inconsistent with the formation of QPO in the inner disc. It has been
suggested that the upper kHz QPO frequency may correspond to a Keplerian frequency in the inner
disc (e.g. Jonker et al. 2000). The removal of the outer parts of the inner disc by 
radiation pressure
but not parts close to the orbital plane is not inconsistent with this, and it may even be that
the disturbance to the disc triggers the oscillation.

%

\subsection{The reduction in blackbody radius}

Church \& Ba\l uci\'nska-Church (2001)
can suggest why the neutron star blackbody emitting area decreases on the
normal and horizontal branches. It is unlikely that this is due to any process on the stellar
surface, and is more likely to be due to processes at the inner disc. In our survey of 
14 LMXBs using {\it ASCA} and {\it BeppoSAX}, 
we investigated the variation of blackbody luminosity with total luminosity. It is well-known
that the blackbody fraction is small in faint sources rising to about 50\% in sources close to 
the Eddington limit. The results were presented geometrically in terms of $h$, the half-height of 
the blackbody emitter, and $H$, the half-height of the inner disc. The blackbody luminosity depends 
on $h$, and is calculated from the blackbody radius (Sect. 3.2.1) while $H$ is proportional to the mass 
accretion rate in the standard theory of the inner, radiatively-supported, disc (Frank et al. 2002). 
It was found that for all sources $h$ $\simeq$ $H$, suggesting that the neutron star emitting area was 
determined by the height of the inner disc. The mechanism for this might be radial flow across the gap 
between disc and neutron star, or alternatively the mechanism of Inogamov \& Sunyaev (1999) which predicts 
that the accretion flow spreads vertically on the star to a height depending on $\dot M$. The survey 
results agreed approximately with this model (Church et al. 2002). 

In a radiatively-supported disc, the vertical height of the disc increases rapidly with radial distance
until at $\sim$10 km from the surface of the neutron star, it has its equilibrium value $H_{\rm eq}$ 
(Frank et al. 2002). In the present work on GX\th 340+0, $H_{\rm eq}$ calculated from 
the source luminosity varies between 30 km at the soft apex and 50 km at the hard 
apex, so that the inner disc towers above the neutron star. If the radiation pressure of the neutron star 
is strong, we would expect some of the inner disc to be blown away above and below the disc, leaving 
a residual inner disc of reduced height. On the basis of the survey results, the height on the neutron star 
would respond to this by
becoming much smaller which may explain $h$ decreasing from $\sim$ 11 km at the soft apex
(where the radiation pressure is small) to $\sim$ 1 km on the horizontal branch.

\subsection{The nature of the Horizontal Branch}

The spectral fitting results show that the luminosity of the Comptonized emission increases 
on the normal branch and then decreases by about the same amount on the horizontal branch,
following the same path in Fig. 5, suggesting that the mass accretion rate first increases, 
then decreases towards the initial value and we might expect the changes on the neutron star
to also reverse at the hard apex. But the neutron star blackbody does not reverse
its behaviour at the hard apex, the temperature and radius changing monotonically
on the NB and HB (Fig. 4). A possible reason that this does not happen could relate to
time delays in the system. An increase in $\dot M$ will affect the emission
of the extended ADC first causing the source to move along the NB towards the hard apex, but changes
in the blackbody emission would be delayed until the increase reaches the
neutron star. Fig. 4 provides evidence for this: the Comptonized emission has increased
before the blackbody changes; Fig. 1 shows that several hours are required for the source
to move halfway along the NB from the soft apex. This timescale is approximately consistent with
estimates of flow times through the disc based on the viscous timescale (Frank et al. 2002).
When $\dot M$ decreases in the disc then $L_{\rm ADC}$ falls, but the increase of $\dot M$ 
at the neutron star could continue for some time. 

We next ask how the source leaves the HB end of the Z-track. Assuming as above that $\dot M$
has already decreased, we would expect the neutron star to cool, radiation pressure to fall 
and $R_{\rm BB}$ to increase. However, the effects in the ADC are not clear. If there was
no change in the emission (as $\dot M$ is constant), the total luminosity would be about constant
and the source may be able to jump from the HB to the NB as the neutron star
cools. However, jumps have not so far been detected, and Z-track sources are seen to move
smoothly from HB to NB.

We can assume that the X-ray luminosity is a good indicator of $\dot M$ when the radiation
pressure is low (on the NB), but this is probably not true when the radiation pressure is
high, so that the decreasing $L_{\rm CPL}$ on the HB may not mean that $\dot M$ falls.
Firstly, the blowing away of outer layers of the inner disc by increased radiation pressure 
may reduce the ADC emission by a geometric factor (giving a reduction in $L_{\rm CPL}$)
while $\dot M$ is constant or rising.
In addition, the radiation pressure can divert a substantial fraction
of $\dot M$ into the jets causing a decrease of $L_{\rm CPL}$ on the HB because of a decrease
of accretion flow in the disc below the corona, reducing the electron density while the total
$\dot M$, including the part diverted to the jets, remains constant, or increases. 
Then the asymmetry between
the blackbody and Comptonized emission disappears, since the reversal of behaviour of
both components takes place at the end of the Z-track, both components reversing their
evolution when the mass accretion rate falls, the source moving back along the Z.



\subsection{The nature of the Flaring Branch}

We have suggested that on the flaring branch there is unstable nuclear burning. This is not
unexpected since the theory of unstable nuclear burning (e.g. Bildsten 1998) shows that 
He burning in a hydrogen-rich environment should be unstable for $\dot M$ $<$ $2.6\times 10^{-8}$
M$_{\sun}$ y$^{-1}$, while for higher $\dot M$, burning is stable.
The Z-track sources have luminosities at about the Eddington limit, but lack of
spectral fitting has often not allowed a definitive statement to be made on whether unstable
burning is expected. 
However, Bildsten (1995) argued that the timescale of flaring in Sco\th X-1
is consistent with the timescale at which unstable burning starting in one part of the
neutron star will spread across the surface. In sources of luminosity mid-way between
burst sources and Z-track sources, we have previously detected bursting and flaring in the 
same observation, as in XB\th 1254-690 (Smale et al. 2002). 

On the flaring branch, the blackbody radius increases systematically 
from its value of 10.8$\pm$1.0 km at the soft apex to $\sim$16.1$\pm$0.8 km. 
If we accept that the whole neutron star is emitting at the soft apex, then in flaring
the emission must extend beyond the surface of the star. Similar values of $R_{\rm BB}$ 
greater than the neutron star radius have been found previously for Z-track sources, such
as a value of 27 km found by Christian \& Swank (1997) for GX\th 5-1.
Lower luminosity X-ray burst sources can 
exhibit radius expansion (Lewin et al. 1985) in which 
$R_{\rm BB}$ increases indicating photospheric expansion due to radiation pressure, 
the luminosity remaining close to the Eddington limit.
In the present work we also have an increase of $R_{\rm BB}$ on the flaring branch which 
we suggest is unstable nuclear burning, and the emitted flux
approaches the Eddington flux. Thus it is possible that radius expansion may take place 
during flaring in the Z-track sources.
The implication is that the emission expands beyond the surface of the neutron star, and the 
deep hole where the height of the inner accretion disc falls 
to that of the neutron star is partly filled by plasma. 
The rise in blackbody temperature is however, only 10\%, and so much less than the
temperature rise in X-ray bursts. 

We next ask whether unstable nuclear burning takes place on all branches
of the Z-track or only on the flaring branch. Conditions on the other branches either prevent 
unstable burning, or 
mask the appearance of flaring. On the flaring branch flares develop from
the lowest X-ray intensity of the source at the soft apex, and are very 
obviously seen in the lightcurve. We have 
examined the lightcurve for variations on timescales
of a few thousand seconds in the other branches, but this was inconclusive.
However, the present results suggest why unstable burning may occur only on the FB. 

At the soft apex, $L_{\rm Tot}$ has its smallest value of $3.7\times 10^{38}$ erg s$^{-1}$, corresponding
to a mass accretion rate per unit area on the neutron star 
$\dot m$ = $1.5\times 10^5$ g cm$^{-2}$ s$^{-1}$. The upper limit for unstable 
nuclear burning is $\dot m$ = $1.3\times 10^5$ g cm$^{-2}$ s$^{-1}$ (Bildsten 1998).
Thus the actual $\dot m$ is close to the theoretical upper limit which Bildsten 
estimates to be accurate to within 30\%, and unstable burning can be possible. 
On the normal branch, the luminosity and mass accretion rate eventually approach a factor of 2 
higher than at the soft apex suggesting that unstable burning is not possible and 
that flaring will not take place on the NB and HB. On the flaring branch, the emitting
blackbody area increases, and so $\dot m$ remains below the critical value, and nuclear
burning remains unstable.
Examination of the 
lightcurve (Fig. 1) shows that flaring begins {\it exactly} at the lowest luminosity of
the source. Thus if we move towards the soft apex along the NB,
it may be that unstable burning begins immediately $\dot m$ falls below the critical value.
Thus the identification of the flaring branch with unstable nuclear burning appears to
be consistent with theory, as is the lack of unstable nuclear burning on the
NB and HB. 


\subsection{The formation of jets by radiation pressure}

Explaining the formation of relativistic jets from the neighbourhood of compact objects
in both Galactic sources and in AGN has been a major astrophysical goal. Two main types
of model have been proposed. First, it has long been suspected that radiation pressure
may play an important role (Begelman \& Rees 1984), and Lynden-Bell (1978) suggested
that a radiatively-thickened inner accretion disc will define two funnels which
allow jet formation perpendicular to the disc, driven by radiation pressure. Second, the other
type of model depends on electromagnetic production via strong electric fields arising from
magnetic fields in the disc around a black hole (Blandford \& Znayek 1977). The present
work reveals the link between jet formation and the position on the Z-track
where the radiation pressure is high, strongly indicating the importance of radiation pressure.

It is well known that radio emission is detected from Z-track sources on the horizontal 
branch, and upper normal branch providing strong evidence that jets are present. On this part of the 
Z-track the high neutron star temperatures we determine mean that the radiation pressure of the 
neutron star is  high, the emitted flux approaching three times super-Eddington (Fig. 7).
It thus appears significant that jets are formed on this part of the Z-track
suggesting that the launching of jets is due to the strong radiation pressure. Our spectral
fitting result that the emitting area on the neutron star decreases is consistent with
matter being blown away from the inner disc, and the release of matter within the system
is directly supported by the measured increase of column density to a maximum at the hard apex.
We thus propose that
jets are formed by the very strong radiation pressure of the neutron star at this part of
the Z-track, where the radiation flux substantially exceeds the Eddington flux. The emission
area is small implying it is from an equatorial strip around the
neutron star, and this geometry favours upwards direction of the jet. Radiation pressure acting horizontally
may disturb the inner disc. However, a line from the equatorial belt to the
inner edge of the disc will rise steeply upwards,
the radiation pressure acting in a direction approaching vertical.

Jets are not seen on the flaring branch, and this is consistent with the model 
proposed. The radiation pressure is not so strong and the blackbody emitter apparently 
expands beyond the neutron star surface. There is normally a conical funnel in the central 
radiatively-supported inner accretion disc (Frank et al. 2002)
where the disc height falls rapidly towards the stellar surface providing an opening for jet 
formation on the NB and HB. However, in flaring this funnel may be partly or completely blocked by plasma
this also opposing jet formation.



\thanks{
This work was supported in part by the Polish KBN grant KBN-1528/P03/2003/25
and by PPARC grant PPA/G/S/2001/00052. We thank the referee for his very useful comments.}


\begin{thebibliography}{}

\bibitem[]{}
Agrawal, V. K., \& Sreekumar, P. 2003, MNRAS, 346, 933

\bibitem[]{}
Ba\l uci\'nska-Church, M., \& Church, M. J. 2005, Proc of ``Interacting Binaries: Accretion, 
Evolution and Outcomes'', Cefalu, July 2004, AIP Conf Proc, New York, 797, 339

\bibitem[]{}
Ba\l uci\'nska-Church, M., Church, M. J., Oosterbroek, T.,
Segreto, A., Morley, R., \& Parmar, A. N. 1999, A\&A, 349, 495

\bibitem[]{}
Ba\l uci\'nska-Church, M., Humphrey, P. J., Church, M. J., \& Parmar, A. N. 2000, A\&A, 360, 583

\bibitem[]{}
Ba\l uci\'nska-Church, M., Barnard, R., Church, M. J., \& Smale, A. P. 2001, A\&A, 378, 847

\bibitem[]{}
Barnard, R., Ba\l uci\'nska-Church, M., Smale, A. P., \& Church, M. J. 2001, A\&A, 380, 494

\bibitem[]{}
Barnard, R., Church, M. J., \& Ba\l uci\'nska-Church, M. 2003, A\&A, 405, 237

\bibitem[]{}
Begelman, M. C., \& Rees, M. J. 1984, MNRAS, 206, 209

\bibitem[]{}
Bildsten, L. 1995, ApJ, 438, 852

\bibitem[]{}
Bildsten, L. 1998, in R. Buccheri, J. van Paradijs \& M. A. Alpar, eds.,
``The Many Faces of Neutron Stars'', Proc NATO ASIC 515,
Dordrecht-Kluwer, 419

\bibitem[]{}
Blandford, R. D., \& Znayek, R. L. 1977, 179, 433

\bibitem[]{}
Bradshaw, C. F., Geldzahler, B. J., \& Fomalont, E. B. 2003, ApJ, 592, 486
%


\bibitem[]{}
Church, M. J. 2001, Adv Space Res, 28, 323

\bibitem[]{}
Church, M. J., \& Ba\l uci\'nska-Church, M. 1995, A\&A, 300, 441

\bibitem[]{}
Church, M. J., \& Ba\l uci\'nska-Church, M. 2001, A\&A, 369, 915

\bibitem[]{}
Church, M. J., \& Ba\l uci\'nska-Church, M. 2004, MNRAS, 348, 955
                                                          
\bibitem[]{}
Church, M. J., Mitsuda, K., Dotani, T., Ba\l uci\'nska-Church, M., 
Inoue, H., \& Yoshida, K. 1997, ApJ, 491, 388
                                                                             
\bibitem[]{}
Church, M. J., Ba\l uci\'nska-Church, M., Dotani, T., \& Asai, K. 1998a, ApJ, 504, 516
                                               
\bibitem[]{}
Church, M. J., Parmar, A. N., Ba\l uci\'nska-Church, M., Oosterbroek, T., Dal Fiume, D.,
\& Orlandini, M. 1998b, A\&A, 338, 556

\bibitem[]{}
Church, M. J., Inogamov, N. A., \& Ba\l uci\'nska-Church, M. 2002, A\&A, 390, 139

\bibitem[]{}
Church, M. J., Reed, D., Dotani, T., \& Ba\l uci\'nska-Church, M., Smale, A. P. 2005, 
359, 1336

\bibitem[]{}
Christian, D. J., \& Swank, J. H. 1997, ApJSupp, 109, 177

\bibitem[]{}
Dickey, J, M., \& Lockman, F. J. 1990, Ann Rev Ast Astr 28, 215

\bibitem[]{}
di Salvo, T., Stella, L. 2002, in A. Goldwurm, D. Neumann, J. Tran Than Van, eds.,
Proc of XXII Moriond astrophysics meeting: ``The Gamma-ray Universe'', Les Arcs,
March 2002, the Gioi Publishers, Vietnam

\bibitem[]{}
di Salvo, T., Stella, L.,  Robba, N. R., et al. 2000, ApJ, 544, L119

\bibitem[]{}
di Salvo, T., Robba, N. R., Iaria, R., Stella, L., Burderi, L., 
\& Israel, G. L. 2001, ApJ, 554, 49

\bibitem[]{}
di Salvo, T., Farinelli, R., Burderi, L., et al. 2002, A\&A, 386, 535

\bibitem[]{}
Done, C., \.Zycki, P., \& Smith, D. A. 2002, MNRAS, 331, 453

\bibitem[]{}
Fender, R. P., \& Hendry, M. A. 2000, MNRAS, 317, 1

\bibitem[]{}
Frank, J., King, A. R., \& Raine, D. J. 2002, ``Accretion Power in Astrophysics'', third edition, 
Cambridge University Press

\bibitem[]{}
Fujimoto, M. Y., Hanawa, T., \& Miyaji, S. 1981, ApJ, 247, 267

\bibitem[]{}
Fushiki, I., \& Lamb, D. Q. 1987, ApJ, 323, L55

\bibitem[]{}
Hasinger, G., \& van der Klis, M. 1989, A\&A, 225, 79

\bibitem[]{}
Hasinger, G., Priedhorsky, W., \& Middleditch, J. 1989, ApJ, 337, 843

\bibitem[]{}
Hasinger, G., van der Klis, M., Ebisawa, K., Dotani, T., \& \hbox{Mitsuda, K.} 1990,
A\&A, 235, 131

\bibitem[]{}
Inogamov, N. A., \& Sunyaev, R. A. 1999, AstL, 25, 269

\bibitem[]{}
Jahoda, K., Swank, J. H., Giles, A. B., et al. 1996, SPIE, 2808, 59

\bibitem[]{}
Jonker, P. G., Wijnands, R., van der Klis, M., Psaltis, D., Kuulkers, E.,
\& Lamb, F. K. 1998, ApJ, 499, L191

\bibitem[]{}
Jonker, P. G., van der Klis, M., Wijnands, R., et al. 2000, ApJ, 537, 374

\bibitem[]{}
Kahn, S. M., \& Grindlay, J. E. 1984, ApJ, 281, 826


\bibitem[]{}
Kuulkers, E., van der Klis, M., Oosterbroek, T., et al. 1994, A\&A, 289, 759

\bibitem[]{}
Kuulkers, E., van der Klis, M.,  van Paradijs, J. 1995, ApJ, 450, 748

\bibitem[]{}
Kuulkers, E., van der Klis, M.,  Oosterbroek, T., van Paradijs, J., 
Lewin, W. H. G. 1997, MNRAS, 287, 495

\bibitem[]{}
Kuulkers, E., Homan, J., van der Klis, M., Lewin, W. H. G., \& Mend\'ez, M. 2002,
A\&A, 382, 947

\bibitem[]{}
Lewin, W. H. G., van Paradijs, J., \& Taam, R. E. 1995, in ``X-ray Binaries'', eds. W. H. G. Lewin,
J. van Paradijs, E. P. J. van den Heuvel, Cambridge University Press.

\bibitem[]{}
Lynden-Bell, D. 1978, Phys Scripta 17, 185


\bibitem[]{} Makishima K. 1986, ``The Physics of Accretion onto Compact Objects'', Proc. of
Tenerife workshop, 1986, Lecture Notes in Physics, 266, Springer-Verlag, p249

\bibitem[]{}
Migliari, S., \& Fender, R. P. 2006, MNRAS, 366, 79

\bibitem[]{}
Mitsuda, K., Inoue, H., Nakamura, N., \& Tanaka, Y. 1989, PASJ, 41, 97

\bibitem[]{}
Penninx, W. 1989, in J. Hunt and B. Battrick, eds., ``Proceedings of the 23rd ESLAB 
Symposium on Two Topics in X-ray Astronomy'', Bologna, Sept. 1989, ESA publications
ESA SP-296, 185

\bibitem[]{}
Penninx, W., Lewin, W. H. G., Tan, J., Mitsuda, K., van der Klis, M., 
\& van Paradijs, J. 1991, MNRAS, 249, 113

\bibitem[]{}
Penninx, W., Zwarthoed, G. A. A., van Paradijs, J., van der Klis, M., Lewin, W. H. G.,
\& Dotani, T. 1993, A\&A, 267, 92

\bibitem[]{}
Penninx, W., Lewin, W. H. G., Tan, J., Mitsuda, K., van der Klis, M., 
\& van Paradijs, J. 1991, MNRAS, 249, 113
\bibitem[]{}
Petrucci, P. O., Haardt, F., Maraschi, L., et al. 2001, ApJ, 556, 716

\bibitem[]{}
Priedhorsky, W., Hasinger, G., Lewin, W. H. G., et al. 1986, ApJ, 306, L91

\bibitem[]{}
Psaltis, D., Lamb, F. K., \& Miller, G. S. 1995, ApJ, 454, L137

\bibitem[]{}
Schatz, H., Bildsten, L., Cumming, A., \& Wiescher, M. 1999, ApJ, 524, 1014

\bibitem[]{}
Schulz, N. S., \& Wijers, R. A. M. J. 1993, A\&A, 273, 123

\bibitem[]{}
Schulz, N. S.,  Hasinger, G., \& Tr\"umper, J. 1989, A\&A, 225, 48

\bibitem[]{}
Shapiro, S. L., Lightman, A. P., Eardley D. M. 1976, ApJ, 204, 187

\bibitem[]{}
Smale, A. P. 1998, ApJ, 498, L141

\bibitem[]{}
Smale, A. P., Church, M. J., \& Ba\l uci\'nska-Church, M. 2001, ApJ, 550, 962

\bibitem[]{}
Smale, A. P., Church, M. J., \& Ba\l uci\'nska-Church, M. 2002, ApJ, 581, 1286

\bibitem[]{}
Sztajno, M., van Paradijs, J., Lewin, W. H. G., Langmeier, A., Tr\"umper, J., 
\& Pietsch, W. 1986, MNRAS, 222, 499

\bibitem[]{}
Tawara, Y., Hirano, T., Kii, T., Matsuoka, M., \& Murakami, T. 1984, PASJ, 36, 861

\bibitem[]{}
Titarchuk, L. 1994, ApJ, 434, 313

\bibitem[]{}
van der Klis, M., Stella, L., White, N. E., Jansen, F., 
\& Parmar, A. N. 1987, ApJ, 316, 411


\bibitem[]{}
van Paradijs, J., Hasinger, G., Lewin, W. H. G., et al. 1988, MNRAS, 231, 379

\bibitem[]{}
van Paradijs, J., Allington-Smith, J., Callanan, P., et al. 1990, A\&A, 235, 156


\bibitem[]{}
Vrtilek, S. D., Raymond, J. C., Garcia, M. R., Verbunt, F., Hasinger, G., 
\& Kurster, M. 1990, A\&A, 235, 162

\bibitem[]{}
Wijnands, R. A. D., van der Klis, M., Kuulkers, E., Asai, K., 
\& Hasinger, G. 1997, A\&A, 323, 399


\bibitem[]{}
White, N. E., Peacock, A., Hasinger, G., et al. 1986, MNRAS, 218, 129


\end{thebibliography}
\end{document}